\documentclass[11pt,a4paper]{article}

\usepackage{mathtools,cancel}

\usepackage{soul,xcolor}
\usepackage{xcolor}
\usepackage{multirow}
\usepackage{pstricks}
\usepackage{dcolumn}
\usepackage{bm}
\usepackage{latexsym}
\usepackage{dcolumn}
\usepackage[utf8x]{inputenc} 
\usepackage{amsmath}
\usepackage{amsfonts,amssymb}
\usepackage{graphicx,epsfig}
\usepackage{color}
\usepackage{psfrag}
\usepackage{amsthm}
\usepackage{ulem} 
\usepackage{soul} 
\usepackage{flushend}
\usepackage{titlesec}
\usepackage{slashed}
\usepackage{authblk}
\usepackage[font=footnotesize]{caption}
\usepackage[hmarginratio=1:1,top=32mm,columnsep=60pt]{geometry}

\newcommand{\bea}{\begin{eqnarray}}
\newcommand{\eea}{\end{eqnarray}}
\newcommand{\be}{\begin{equation}}
\newcommand{\ee}{\end{equation}}
\newcommand{\ba}{\begin{array}}
\newcommand{\ea}{\end{array}}

\begin{document}

\unitlength = 1mm
\begin{flushright}
\end{flushright}

\title{\textbf{Searching for neutral state in the rare decay $J/\psi \rightarrow e^+ e^- \phi$}}

\author{\normalsize Ahmed Rashed}
\affil{\small
Department  of Physics,  Shippensburg University of Pennsylvania,\\
 Franklin Science Center, 1871 Old Main Drive, Pennsylvania, 17257, USA
 }
\date{}

{\let\newpage\relax\maketitle}

\begin{abstract}
\noindent \normalsize 

We investigate the decay process $J/\psi \rightarrow e^+ e^-\phi$, where the relatively clean electromagnetic (EM) transitions dominate at leading order at the tree level, while hadronic contributions arise only through hadronic loop transitions. The branching ratio of $J/\psi \rightarrow e^+ e^-\phi$ was estimated to be approximately $2.28 \times 10^{-8}$ keV where the hadronic effects are negligible compared to the EM contributions. The upper limit on the branching fraction was set to be $B(J/\psi \rightarrow e^+ e^-\phi) < 1.2 \times 10^{-7}$ by BESIII collaboration. In this paper we investigate the existence of a neutral state such as dark photon and dark $Z$. Our analysis shows that the constraints on the dark photon and dark $Z$ are stringent in the way that large enhancement to the SM branching ratio up to the present experimental limit is not possible. Therefore, observing a signal for the dark photon and dark $Z$ in this decay channel is changeable. 
\end{abstract}


\newpage

\section{Introduction}

High-intensity experiments enable precise tests of the Standard Model (SM) and the search for signs of new physics beyond the Standard Model (BSM). Rare decays are particularly useful in the low-energy regime, as SM contributions are suppressed, making anomalies potential indicators of BSM effects. Experiments like BEPCII/BESIII \cite{BESIII:2012pbg} and BELLE-II \cite{Belle:2006jvm, Belle:2007dxy} offer extensive data for studying such rare processes.

The decay $J/\psi \rightarrow e^+ e^-\phi$ serves as an example of a rare process measurable by BESIII. The dominant contribution comes from electromagnetic (EM) transitions via vector meson dominance \cite{Guberina:1980dc}, while sub-leading contributions arise from hadronic meson loops involving quark-gluon dynamics. Experimental data on radiative decays $J/\psi \rightarrow \gamma P$ and $\gamma S$, $\phi \rightarrow \gamma P$ (where $P$ and $S$ are pseudoscalar and scalar mesons) helps estimate hadronic contributions.

Results indicate hadronic effects are smaller than EM contributions, with significant intermediate states being $\eta, \eta^\prime$ (pseudoscalar) and $f_0 (980)$ (scalar), while contributions from $\sigma(500)$ can be neglected due to weak couplings. This analysis provides insights into rare processes and potential traces of BSM physics.

Flavor-changing neutral currents (FCNCs) are powerful tools for exploring new physics (NP) and place strong constraints on extensions of the Standard Model (SM). This study focuses on FCNC processes in \( B \) and \( K \) decays to investigate models involving a light gauge boson. Specifically, it examines the dark \( U(1)_D \) model, which predicts a dark photon or a dark \( Z \) (\( Z' \)), collectively referred to as \( Z_D \) \cite{Holdom:1985ag, Gopalakrishna:2008dv, Davoudiasl:2012ag}. 

The nature of \( Z_D \) depends on its interactions: kinetic mixing with the electromagnetic field produces a dark photon with vector couplings to SM fermions (except neutrinos), while mass mixing with SM gauge fields results in a dark \( Z \). FCNC processes involving \( Z_D \) are loop-mediated, with significant effects in \( B \) decays due to the suppression of the GIM mechanism for up-type quarks, while \( D \) decays are less affected due to down-type quark loops.

Unlike heavy mediators, light mediators like \( Z_D \) lead to \( q^2 \)-dependent Wilson coefficients. This paper evaluates \( Z_D \)-mediated \( b \to s \ell^+ \ell^- \) transitions for \( Z_D \) masses in the range \( 0.01 \, \text{GeV} \lesssim M_{Z_D} \lesssim 2 \, \text{GeV} \) \cite{Datta:2017pfz, Sala:2017ihs, Bishara:2017pje, Ghosh:2017ber, Datta:2017ezo, Altmannshofer:2017bsz, Datta:2018xty, Datta:2019zca,Darme:2020hpo,Borah:2020swo, Darme:2021qzw,Crivellin:2022obd}, considering both on-shell and off-shell decays. Hadronic decays and invisible decays of \( Z_D \) (from interactions with dark sector particles) are included. Additionally, the study explores models where \( Z_D \) directly couples to muons and/or electrons, providing a detailed analysis of the rates and effects of these processes.

In this paper, we will be interested in probing dark photon and dark $Z$ states via the decay channel $J/\psi \rightarrow e^+ e^-\phi$. Constraints on the model parameters are taken from $B_s$ mixing, $B_s \to \mu^+ \mu^-$, $B \to K^{(*)} \nu \bar{\nu}$, Kaon decay and mixing, Radiative $K^+ \to \mu^+ \nu_\mu Z_D$ decays, Radiative $\pi^+ \to \mu^+ \nu_\mu Z_D$ decays, Atomic parity violation, Neutrino trident and CE$\nu$NS, and Collider and other bounds. All these constraints were discussed in Ref. 

The paper is organized as follows: In Sec. 2, the general formalism of the dark symmetry $U(1)_D$ is discuss. In Sec. 3, we discuss the SM formalism and numerics to the decay process  $J/\psi \rightarrow e^+ e^-\phi$. The NP contribution is discussed in Sec. 4 and we summarize in Sec. 5.  

\section{Formalism}

We consider \( Z_D \) as a gauge boson associated with the broken \( U(1)_D \) symmetry of a dark sector, coupled to the Standard Model (SM) via kinetic mixing with \( U(1)_Y \) \cite{Holdom:1985ag}. The gauge Lagrangian is given by \cite{Davoudiasl:2012ag}:

\[
\mathcal{L}_\text{gauge} = -\frac{1}{4} B_{\mu\nu} B^{\mu\nu} + \frac{1}{2} \frac{\varepsilon}{\cos \theta_W} B_{\mu\nu} Z_D^{\mu\nu} - \frac{1}{4} Z_{D \mu\nu} Z_D^{\mu\nu},
\]
where \( B_{\mu\nu} \) and \( Z_{D\mu\nu} \) are the field strength tensors for \( B \) and \( Z_D \), \( \varepsilon \) is the kinetic mixing parameter, and \( \theta_W \) is the weak mixing angle.

After diagonalizing the gauge sector \cite{Chun:2010ve, Davoudiasl:2012ag}, \( Z_D \) acquires an induced coupling to the SM electromagnetic current, leading to the *dark photon* model. To leading order in \( \varepsilon \), the interaction is:

\[
\mathcal{L}_D^\text{em} \supset e \varepsilon Z_D^\mu J_\mu^\text{em} - i e \varepsilon \left[\left[Z_D W^+ W^-\right]\right],
\]
where \( \left[\left[Z_D W^+ W^-\right]\right] \) represents the interaction of \( Z_D \) with \( W^\pm \) bosons.
\begin{equation}
\left[\left[Z_D W^+ W^- \right]\right] = \varepsilon^\mu_{Z_D}(k_1) \varepsilon^\nu_{W^+}(k_2) \varepsilon^\lambda_{W^-}(k_3) 
 \times\left[ (k_1 - k_2)_\lambda g_{\mu\nu} + (k_2 - k_3)_\mu g_{\nu\lambda} + (k_3 - k_1)_\nu g_{\lambda\mu} \right]\,.
\end{equation}

If \( U(1)_D \) is broken by a scalar field charged under the SM, \( Z_D \) can mix with the SM \( Z \) boson via mass terms \cite{Gopalakrishna:2008dv, Davoudiasl:2012ag}. The physical states are expressed as:

\[
Z = Z^0 \cos\xi - Z_D^0 \sin\xi, \quad Z_D = Z^0 \sin\xi + Z_D^0 \cos\xi,
\]
where \( \xi \) is the mass mixing angle. This interaction defines the *dark \( Z \)* model, with the Lagrangian:

\[
\mathcal{L}_D^Z \supset \frac{g}{\cos \theta_W} \varepsilon_Z Z_D^\mu J_\mu^Z - i g \cos \theta_W \varepsilon_Z \left[\left[Z_D W^+ W^-\right]\right],
\]
where \( \varepsilon_Z = \frac{1}{2} \tan 2\xi \). If \( U(1)_D \) is broken by SM singlet scalars, then \( \varepsilon_Z = 0 \), reducing the model to a dark photon.

We do not specify a Higgs sector for \( U(1)_D \) breaking but focus on the general mass mixing. The free parameters are the mixing parameters \( \varepsilon \) and \( \varepsilon_Z \), along with the mass of \( Z_D \) (\( M_{Z_D} \)). Updated constraints on these parameters are provided in recent studies.

In the relevant mass range, \( Z_D \) decays into lepton pairs and hadronic final states. While \( e^+e^- \) and \( \nu\bar{\nu} \) decays are always kinematically allowed, the \( \mu^+ \mu^- \) channel is accessible only if \( M_{Z_D} > 2m_\mu \). The decay widths for these processes are given by:

\[
\Gamma(Z_D \rightarrow \ell^+ \ell^-) = \frac{e^2}{96 \pi c_W^2 s_W^2 M_{Z_D}} \sqrt{1 - 4\frac{m_\ell^2}{M_{Z_D}^2}} \Bigg[8 c_W^2 s_W^2 \varepsilon^2 (2m_\ell^2 + M_{Z_D}^2)
\]
\[
- 4 c_W s_W (4s_W^2 - 1) \varepsilon \varepsilon_Z (2m_\ell^2 + M_{Z_D}^2) 
+ \varepsilon_Z^2 \big(m_\ell^2 (16s_W^4 - 8s_W^2 - 1) + M_{Z_D}^2 (8s_W^4 - 4s_W^2 + 1)\big)\Bigg],
\]

\[
\Gamma(Z_D \rightarrow \nu \bar{\nu}) = \frac{e^2 M_{Z_D} \varepsilon_Z^2}{96 \pi c_W^2 s_W^2}.
\]

Hadronic decays, however, cannot be calculated directly using perturbative QCD. Instead, the vector meson dominance (VMD) model is employed to describe low-energy QCD \cite{Sakurai:1960ju, Kroll:1967it, Lee:1967iv, Fraas:1969uwg, Bando:1984ej}. Recent data-driven studies have calculated the hadronic decay widths of light \( U(1) \) vector bosons \cite{Tulin:2014tya, Ilten:2018crw, Foguel:2022ppx}. For dark photons with couplings proportional to electric charge, the hadronic decay width can be expressed as:

\[
\Gamma(Z_D \to \mathcal{H}) = \Gamma(Z_D \to \mu^+ \mu^-) \times \mathcal{R}^\mathcal{H}_\mu,
\]

where \( \mathcal{R}^\mathcal{H}_\mu = \sigma(e^+e^- \to \mathcal{H}) / \sigma(e^+e^- \to \mu^+ \mu^-) \), measured experimentally \cite{ParticleDataGroup:2022pth}. At energies away from hadron resonances, \( e^+e^- \) annihilation gradually transitions to perturbative quark-pair production, with \( \mathcal{R}^\mathcal{H}_\mu \simeq N_c \sum_{f=u,d,s} (Q_{em}^f)^2 = 2 \), where \( N_c = 3 \) is the color factor and \( Q_{em}^f \) is the fermion's electric charge.

For baryophilic dark photons, the hadronic decay width has been calculated by summing over various final states within the VMD framework \cite{Foguel:2022ppx}. This analysis is adapted to the vector coupling of the dark \( Z \). For axial vector couplings, quark-level decays are used to estimate contributions to the hadronic width. 

\section{The EM Contribution to $J/\psi\rightarrow e^+e^-  \phi$}

\begin{figure}[!h]
\centering
\begin{minipage}[!htbp]{0.6\textwidth}
\centering
\includegraphics[width=0.98\textwidth]{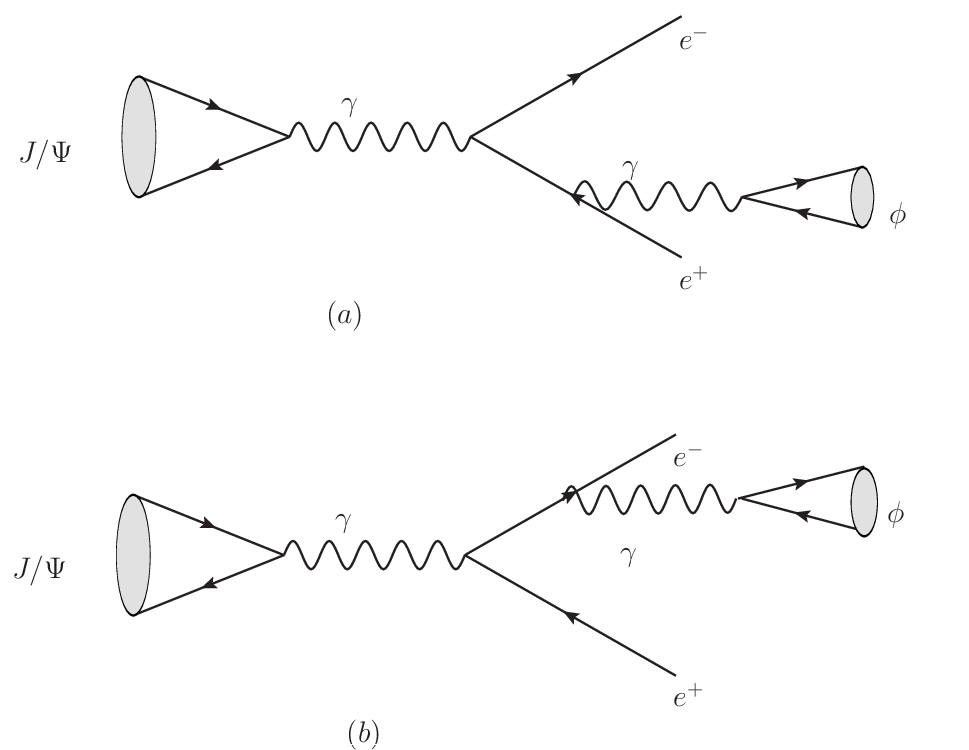}
\caption{The EM transitions of $J/\psi\to e^+e^-\phi$.}
\label{qed}
\end{minipage}
\end{figure}

The leading EM contribution to $J/\psi\rightarrow e^+e^-\phi$ for which the corresponding Feynman diagrams
are shown in Fig.~\ref{qed} (Ref.~\cite{Guo:2015bqi})

The decay \( J/\psi \to \gamma \phi \) is forbidden by \( C \)-parity conservation unless new physics beyond the Standard Model (BSM) exists to break this rule. However, when the photon is virtual and converts into an \( e^+e^- \) pair, the decay \( J/\psi \to e^+e^-\phi \) becomes allowed and measurable. 

\textbf{Electromagnetic (EM) Contribution:} The leading contribution is electromagnetic (EM), as shown in Fig.~\ref{qed}. The effective \( \gamma \)-vector meson couplings are calculated using vector meson dominance (VMD):
\[
\langle J/\psi|\bar{c}\gamma^\nu c|0\rangle = g_{\psi\gamma}\varepsilon^{*\nu}_\psi, \quad
\langle\phi|\bar{u}\gamma^\mu u|0\rangle = g_{\phi\gamma}\varepsilon^{*\mu}_\phi,
\]
where the couplings are
\[
g_{\psi\gamma} = 0.150~\text{GeV}^2, \quad g_{\phi\gamma} = 0.013~\text{GeV}^2.
\]
The EM transition amplitude is
\[
\mathcal{M}_{EM} = \bar{u}(p_1)\left[(-ie\gamma^\nu)\frac{\rlap /p_2+\rlap /p_\phi+m_e}{(p_2+p_\phi)^2-m^2_e}(-ie\gamma^\mu) 
+ (-ie\gamma^\mu)\frac{\rlap /p_1+\rlap /p_\phi+m_e}{(p_1+p_\phi)^2-m^2_e}(-ie\gamma^\nu)\right]v(p_2)
\frac{g_{\phi\gamma}\varepsilon^{\mu}_\phi}{p_\phi^2}\frac{g_{\psi\gamma}\varepsilon^{*\nu}_\psi}{p_\psi^2}.
\]

\textbf{Hadronic Loop Contributions:} Hadronic contributions via intermediate states (\( f_0(980) \), \( \eta \), \( \eta' \)) are included. The effective couplings are:
\[
g_{\psi f_0} = 3.21 \times 10^{-5}~\text{GeV}, \quad g_{\phi f_0} = 3.03 \times 10^{-2}~\text{GeV}^{-1},
\]
\[
g_{\psi \eta} = 5.36 \times 10^{-4}~\text{GeV}^{-1}, \quad g_{\psi \eta'} = 1.28 \times 10^{-3}~\text{GeV}^{-1}.
\]

\textbf{Decay Widths and Branching Ratios:} The decay widths and branching ratios of \( J/\psi \to e^+e^- \phi \) from EM and hadronic processes are summarized in Table~\ref{decaywidth}:

\begin{table}[htbp]
\caption{The decay widths and branching ratios of $J/\psi\rightarrow e^+e^- \phi$ contributed from the EM and hadronic processes.
\label{decaywidth}}
\begin{center}
\begin{tabular}[c]{|c|c|c|c|c|c|}\hline
                                   &EM        & via $f_0(980)$     &via $\eta$ &via $\eta'$ & total  \\\hline
$\Gamma(J/\psi\rightarrow e^+e^- \phi)$(keV)&$2.12\times10^{-6}$ &$2.00\times10^{-13}$ &$7.2\times10^{-11}$ &$1.12\times10^{-9}$ &$2.12\times10^{-6}$\\\hline
B.R. & $2.28\times 10^{-8}$ & $2.16\times 10^{-15}$ & $7.75\times 10^{-13}$ & $1.20\times 10^{-11}$ & $2.28\times 10^{-8}$ \\\hline
\end{tabular}
\end{center}
\end{table}

The EM process dominates, with a partial branching ratio of \( 2.28 \times 10^{-8} \) keV. Hadronic loop contributions are suppressed by at least three orders of magnitude. The SM background is minimal, making this process a promising probe for BSM physics. The decay \( J/\psi \to e^+e^- \phi \) is a rare process in the SM, dominated by the EM contribution. Its suppression makes it an ideal candidate to explore potential BSM contributions.

\section{Numerical Analysis}

The branching ratio formula is given by


\begin{equation}
BR_{A/Z_D} = \frac{1}{\Gamma} \dfrac{1}{(2 pi)^3 32 M_{J/\psi}^3} \int_{(m_{12}^2)_{\text{min}}}^{(m_{12}^2)_{\text{max}}} \int_{(m_{23}^2)_{\text{min}}}^{(m_{23}^2)_{\text{max}}} |\mathcal{M}_{A/Z_D}+\mathcal{M}_{EM}|^2 d m_{23}^2 d m_{12}^2
\end{equation}

where the dark photon contribution to the decay process \( J/\psi \to e^+e^- \phi \) is

\[
\mathcal{M}_{A} = \bar{u}(p_1)\left[(-ie\epsilon\gamma^\nu)\frac{\rlap /p_2+\rlap /p_\phi+m_e}{(p_2+p_\phi)^2-m^2_e}(-ie\epsilon\gamma^\mu) 
+ (-ie\epsilon\gamma^\mu)\frac{\rlap /p_1+\rlap /p_\phi+m_e}{(p_1+p_\phi)^2-m^2_e}(-ie\epsilon\gamma^\nu)\right]v(p_2)
\frac{g_{\phi\gamma}\varepsilon^{\mu}_\phi}{p_\phi^2}\frac{g_{\psi\gamma}\varepsilon^{*\nu}_\psi}{p_\psi^2}.
\]

and the dark $Z_D$ contribution is 
\begin{eqnarray}
\mathcal{M}_{Z_D} &= & (g_L + g_R) \,  \, \frac{g_{\psi \gamma}}{e} \, \epsilon_Z 
    \, P_{\psi} \, (\bar{\varepsilon}^*)^{\nu^\prime} \,  \, (g_L + g_R) 
    \, \frac{g_{\phi \gamma}}{e} \, \epsilon_Z 
    \, P_{\phi} \, \bar{\varepsilon}^{\mu^\prime} \nonumber \\
& \times & \frac{i \left( \frac{\overline{P_{\psi}}^{\nu} \overline{P_{\psi}}^{\nu^\prime}}{\mathrm{M}_x^2} - \bar{g}^{\nu \nu^\prime} \right)}
    {\overline{P_{\psi}}^2 + i \Gamma_x \mathrm{M}_x^2 - \mathrm{M}_x^2} 
    \, \frac{g}{2 \, \mathrm{C}_W} \, \, \bar{u}(P_1, \mathrm{m}_e) \nonumber \\
& \times & \Bigg( 
    \mathrm{V}^\mu \, \epsilon_Z 
    \, \frac{\bar{\gamma} \cdot \overline{P_1} + \bar{\gamma} \cdot \overline{P_{\phi}} + \mathrm{m}_e}
    {\big(\overline{P_1} + \overline{P_{\phi}}\big)^2 - \mathrm{m}_e^2} 
    \, \mathrm{V}^\nu \, \epsilon_Z \nonumber \\
& + & \mathrm{V}^\nu \, \epsilon_Z 
    \, \frac{\bar{\gamma} \cdot \overline{P_2} + \bar{\gamma} \cdot \overline{P_{\phi}} + \mathrm{m}_e}
    {\big(\overline{P_2} + \overline{P_{\phi}}\big)^2 - \mathrm{m}_e^2} 
    \, \mathrm{V}^\mu \, \epsilon_Z 
    \Bigg) \, v(P_2, \mathrm{m}_e) \nonumber \\
& \times & \frac{i \left( \frac{\overline{P_{\phi}}^{\mu} \overline{P_{\phi}}^{\mu^\prime}}{\mathrm{M}_x^2} - \bar{g}^{\mu \mu^\prime} \right)}
    {\overline{P_{\phi}}^2 + i \Gamma_x \mathrm{M}_x^2 - \mathrm{M}_x^2} 
    \, \frac{g}{2 \, \mathrm{C}_W}.
\end{eqnarray}


with

\begin{eqnarray}
&& V^\mu = \frac{1}{2}\frac{g}{c_W} (g_L \gamma^\mu(1-\gamma^5)+g_R \gamma^\mu(1+\gamma^5)),\\
&&\text{E}_2(\text{m$_{12}^2$})\text{=}\frac{\sqrt{\text{m}_{12}^2}}{2 },\\
&&\text{E}_3(\text{m$_{12}^2$})\text{=}\frac{-\text{m}_{12}^2+\text{M$_{J/\psi} $}^2-\text{M$_\phi $}^2}{2 \sqrt{\text{m}_{12}^2}}\\
&&(\text{m}_{23}^2(\text{m$_{12}^2$}))_\text{(max)}\text{=}(\text{E}_2(\text{m}_{12}^2)+\text{E}_3(\text{m}_{12}^2))^2-\left(\sqrt{\text{E}_2(\text{m}_{12}^2)^2-\text{m}_e^2}-\sqrt{\text{E}_3(\text{m}_{12}^2)^2-\text{M$_\phi $}^2}\right)^2\\
&&(\text{m}_{23}^2(\text{m$_{12}^2$}))_\text{(min)}\text{=}(\text{E}_2(\text{m}_{12}^2)+\text{E}_3(\text{m}_{12}^2))^2-\left(\sqrt{\text{E}_2(\text{m}_{12}^2)^2-\text{m}_e^2}-\sqrt{\text{E}_3(\text{m}_{12}^2)^2+\text{M$_\phi $}^2}\right)^2\\
&&(\text{m}_{12})_\text{(max)}=(\text{M$_{J/\psi} $}-\text{M$_\phi $})^2\\
&&(\text{m}_{12}^2)_\text{(min)}=4\text{m}_e^2
\end{eqnarray}

\begin{figure}[!h]
\centering
\includegraphics[width=0.3\textwidth]{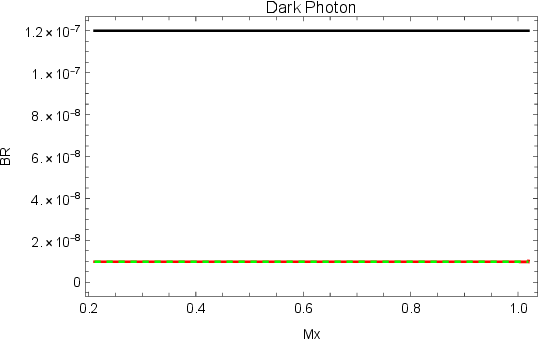}~~
\includegraphics[width=0.3\textwidth]{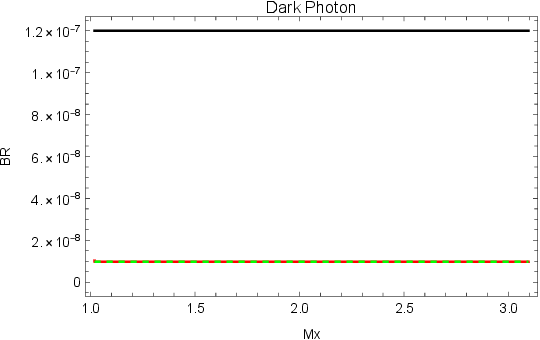}~~
\includegraphics[width=0.3\textwidth]{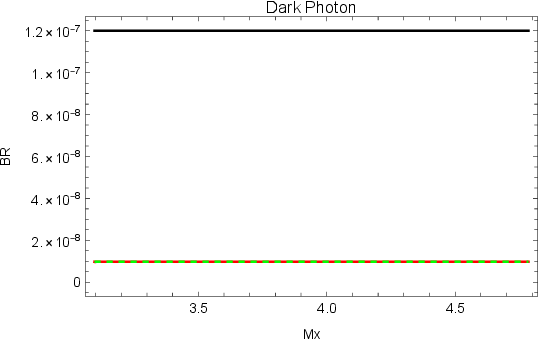}
\caption{The contribution of dark photon to the BR of $J/\psi\to e^+e^-\phi$ with $\varepsilon = 2\times 10^{-2}$ and $\varepsilon^\prime = 0$. The black line represents the experimental limit, the green line represents the SM contribution, and the red line represents the total SM+NP branching ratio.}
\label{qed1}
\end{figure}

\begin{figure}[!h]
\centering
\includegraphics[width=0.3\textwidth]{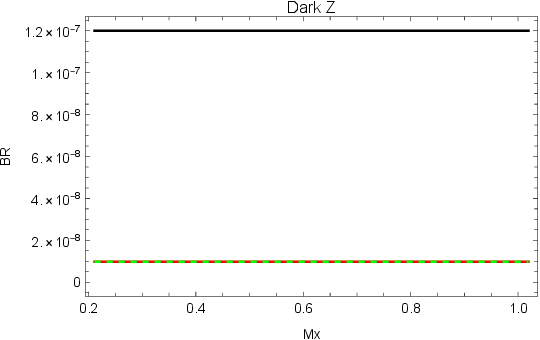}~~
\includegraphics[width=0.3\textwidth]{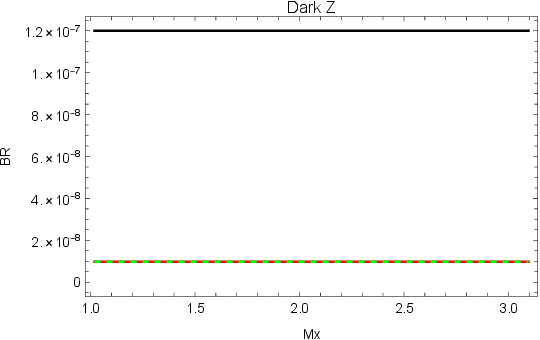}~~
\includegraphics[width=0.3\textwidth]{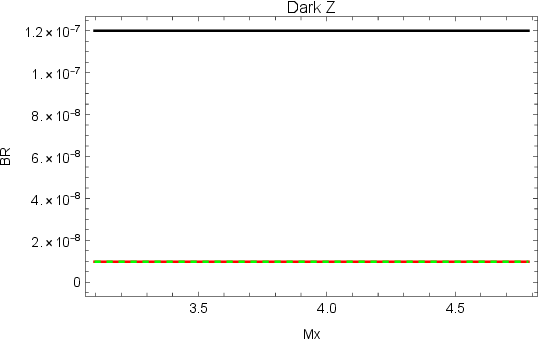}
\caption{The contribution of dark $Z_D$ to the BR of $J/\psi\to e^+e^-\phi$ with $\varepsilon^\prime = 10^{-5}$ and  with $\varepsilon = 0$. The black line represents the experimental limit, the green line represents the SM contribution, and the red line represents the total SM+NP branching ratio.}
\label{qed2}
\end{figure}

In Fig.~\ref{qed1} and Fig.~\ref{qed2} we show the contribution of the dark photn and dark $Z_D$ to the decay channel \( J/\psi \to e^+e^- \phi \) with $\varepsilon = 2\times 10^{-2}$ and $\varepsilon^\prime = 0$ for the first figure and $\varepsilon^\prime = 10^{-5}$ and  with $\varepsilon = 0$ for the second figure. The $M_X$ value taken in the range of $M_{B}-M_{K}$ and $2m_\mu$ as the constraints on the model parameters discussed with this range in Ref.~\cite{Datta:2022zng}. In this range, we avoid the $J/\psi$ and $\phi$ masses to avoid the mixing effect. The contribution is almost negligible because of the stringent constraints on the model parameters. The NP contribution is independent of the new state mass as the Lorentz structure of the vertex is similar to the EM contribution and the decay width of the NP states is way smaller than the mass.

From the graphs above, we show that the dark photon with $\varepsilon = 2\times 10^{-2}$ and dark $Z_D$ with $\varepsilon^\prime = 10^{-5}$ contributions are very small under the stringent constraints. Therefore, probing down the two states of dark photon and dark $Z_D$ in the decay channel $J/\psi\to e^+e^-\phi$ is not possible because of the strong constraints on the model parameters.

\section{Conclusion}

In conclusion, we have analyzed the decay process \( J/\psi \to e^+e^-\phi \), where electromagnetic (EM) transitions dominate at leading order with negligible contributions from hadronic loops. The branching ratio was estimated to be approximately \( 2.28 \times 10^{-8} \), consistent with the dominance of EM effects. The BESIII collaboration has set an experimental upper limit of \( B(J/\psi \to e^+e^-\phi) < 1.2 \times 10^{-7} \). We explored the potential existence of new physics contributions, such as a dark photon or a dark \( Z \), in this decay channel. Our findings indicate that constraints on these hypothetical particles are stringent, making significant enhancements to the SM branching ratio within the current experimental limit unlikely. Therefore, while this decay channel remains a useful probe for new physics, detecting signals for the dark photon or dark \( Z \) in this context appears challenging.

\section*{Appendix: Kinematics}

In this paper we calculate the decay process in Fig.~\ref{qed} in the rest frame of the decaying particle $J/\psi$ where it is momentum is give by $P=(M_{J/\psi},0,0,0)$, where for the other particles are $P_1=P_{e_1^-}$, $P_2=P_{e_1^+}$, and $P_3=P_{\phi}$. The scalar multiplication of the momenta are given as follows
\begin{eqnarray}
P_1 \cdot P_2 &=& \frac{1}{2}(m_{12}^2-2m_e^2),\nonumber\\
P_2 \cdot P_3 &=& \frac{1}{2}(m_{23}^2-m_e^2-m_\phi^2),\nonumber\\
P_1 \cdot P_3 &=& \frac{1}{2}(M_{J/\psi}^2-m_{12}^2-m_{23}^2+m_e^2),\nonumber\\
P \cdot P_1 &=& P_1^2 + P_1 \cdot P_2 + P_1 \cdot P_3,\nonumber\\
P \cdot P_2 &=& P_2^2 + P_1 \cdot P_2 + P_2 \cdot P_3,\; \text{and}\nonumber\\
P \cdot P_3 &=& P_3^2 + P_1 \cdot P_3 + P_2 \cdot P_3,\nonumber\\
\end{eqnarray}
where $P^2=M_{J/\psi}$, $P_1=m_e^2$, $P_2=m_e^2$, and $P_3=m_\phi^2$. $m_{12}^2=(P_1 + P_2)^2=(P-P3)^2$, $m_{23}^2=(P_2 + P_3)^2=(P-P_1)^2$, and $m_{13}^2=(P_1 + P_3)^2=(P-P_2)^2$. Also, $m_{12}^2 + m_{23}^2 + m_{13}^2=M_{J/\psi}^2 + m_1^2 + m_2^2 + m_3^2$. The kinematical equations of the energy and momentum of this frame of reference are
\begin{eqnarray}
E_1 &=& \frac{M_{J/\psi}^2+m_1^2-m_{23}^2}{2 M_{J/\psi}},\nonumber\\
E_2 &=& \frac{M_{J/\psi}^2+m_2^2-m_{13}^2}{2 M_{J/\psi}},\nonumber\\
E_3 &=& \frac{M_{J/\psi}^2+m_3^2-m_{12}^2}{2 M_{J/\psi}},\nonumber\\
P_1 &=& \frac{\lambda^{1/2}(s,m_1^2,m_{23}^2)}{2 M_{J/\psi}},\nonumber\\
P_2 &=& \frac{\lambda^{1/2}(s,m_2^2,m_{13}^2)}{2 M_{J/\psi}},\nonumber\\
P_3 &=& \frac{\lambda^{1/2}(s,m_3^2,m_{12}^2)}{2 M_{J/\psi}},
\end{eqnarray}
where $\lambda(x,y,z)=x^2+y^2+z^2-2xy-2yz-2xz$.
The differential decay rate will be calculated using this equation
\begin{equation}
d\Gamma=\frac{1}{(2\pi)^3}\frac{1}{32M_{H_1}^3}\bar{|M|^2} dm_{12}^2dm_{23}^2,
\end{equation}\\
with
\begin{equation}
\bar{|M|^2}=\sum_{spin}|M_{SM}+M_{NP}|^2=\sum_{spin}|M_{\gamma}+M_{Z}+M_{X}|^2,
\end{equation}
The limits of the kinematical variables.
\begin{eqnarray}
&&(m_1+m_2)^2 \leq m_{12}^2 \leq (M_{J/\psi}-m_3)^2,\nonumber\\
&&(m_2+m_3)^2 \leq m_{23}^2 \leq (M_{J/\psi}-m_1)^2,\nonumber\\
&&(m_1+m_3)^2 \leq m_{13}^2 \leq (M_{J/\psi}-m_2)^2.
\end{eqnarray}




\bibliographystyle{JHEP}
\bibliography{ref}

\end{document}